\documentclass[conference]{IEEEtran}

\usepackage{multirow}
\usepackage{amsfonts}
\usepackage{amsmath}
\usepackage{color}
\usepackage{bm}
\usepackage{amssymb}
\usepackage{graphicx}
\usepackage{algorithmicx}
\usepackage{algpseudocode}
\usepackage{algorithm}
\usepackage{fancyhdr}
\usepackage{epstopdf}
\usepackage{paralist}
\usepackage{url}
\usepackage{subfigure}
\usepackage[final]{changes}
\usepackage[english]{babel}
\IEEEoverridecommandlockouts

\makeindex

\begin{document}

\title{IoT Cloud-based Distribution System State Estimation: Virtual Objects and Context-Awareness}

\author{\IEEEauthorblockN{Alessio Meloni, Paolo Attilio Pegoraro, Luigi Atzori, Paolo Castello, Sara Sulis}
\IEEEauthorblockA{DIEE, University of Cagliari, Italy\\
 \textit{\{alessio.meloni,paolo.pegoraro,l.atzori,paolo.castello,sara.sulis\}@diee.unica.it}\\
}

\thanks{This work has been supported by Regione Autonoma della Sardegna, L.R. 7/2007: Promozione della ricerca scientifica e dell'innovazione tecnologica in Sardegna, annualit\`{a} 2012, CRP-60511".  \copyright 2016 IEEE. The IEEE copyright notice applies.}

}

\maketitle

\begin{abstract}
This paper presents an IoT cloud-based state estimation system for distribution networks in which the PMUs 
(Phasor Measurement Units) are virtualized with respect to the physical devices. In the considered system only application level entities 
are put in the cloud, whereas virtualized PMUs are running in the communication network edge (i.e. closer to the physical objects) in 
order to \replaced{have}{implement} a certain degree of local logic, which allows to \deleted{realize}\added{implement} a bandwidth-efficient and smart 
data transmission to the involved applications in the cloud. The major contributions of the paper are the following: we demonstrate that a 
cloud-based architecture is capable of achieving the QoS level required by the specific state estimation application; we show that 
implementing a certain local logic for data transmission in the cloud, the result of the state estimation is not degraded with respect to 
the case of an estimation that takes place frequently at fixed intervals; we show the results in terms of latency and reduced network load 
for a reference smart grid network.\deleted{In addition, estimation results in terms of rms voltage values are presented. They prove the 
capability of the proposed system to follow possible dynamics.}
\end{abstract}

\begin{IEEEkeywords}
\added{Distribution System} State Estimation, Internet of Things, Phasor Measurement Units, Cloud, Virtualization. \\
\end{IEEEkeywords}

\vspace{-0.5cm}

\section{Introduction}
The metamorphosis of power distribution systems from \deleted{pure electrical} \added{electric} grids into smart grids is currently underway and it is consistently changing, among the others, how operational tasks are performed \cite{Heydt_NextGenDistrSyst}. In particular, the introduction of Distributed Energy Resources (DERs) 
 requires Distribution System Operators (DSO) to rethink the way power systems are managed so as to face 
uncertainty \added{sources} and quick dynamics that were previously not present \cite{MuscasTIM2015}. In this sense, the increasingly wide deployment of sensing 
devices such as Phasor Measurement Units (PMUs) and Smart Meters (SMs) can be seen as the means to estimate the state of the network in a 
more efficient\deleted{, dynamic and quick } \added{and dynamic} way, in order \deleted{promptly react}\added{to react promptly} in case operational actions must be taken.

In light of these fast-pace changes in the \deleted{electrical}\added{electric} grid domain, also the underlying requirements of the associated information and 
communication network have significantly changed, thus calling for a modernization and sometimes creation of a new ICT infrastructure 
which should be able to keep up with the needs of smart grid operators \cite{Gungor2013}. As an example, PMUs are able to sense and send 
data with frequencies as high as 50 times per second (or 60 times per second for 60 Hz electric systems). This gives the possibility to 
have fine-grained 
information about the state of a node of the \deleted{electrical}\added{electric} network, but it poses several questions on where \deleted{all these }data should be sent 
(remotely or not), if it is really necessary to send all this data and at which reporting rate, when this should be sent and how to make 
this data available for use by multiple applications. Considering these premises, the following characteristics are foreseen for the 
future ICT infrastructure considering such devices: a scalable and elastic infrastructure capable of accomodating big data flows and their 
related storage and computation operations; an infrastructure which ensures flexibility by giving the possibility to make decisions about 
data transmission in a context-aware manner; the possibility to virtualize (i.e. abstract from the specific language of the physical 
devices) data sensed in order to make them available for multiple applications. 

On the basis of these requirements, the cloud computing technologies\added{, see for example \cite{Bera2015} and \cite{Fang2012},} can be seen as an enabling information and communication technology for smart grid operation's applications such as Distribution System State Estimation (DSSE). Cloud-based solutions can solve the non-trivial tasks related to storage, real-time computation and optimization of a large amount of data as that generated in a complex system of systems as the smart grid is. In the last years, some works on the use of a cloud-based model for distribution networks have been proposed. In \cite{Rusitschka2010}, a model for smart grid data management based on cloud computing was proposed. Among the use cases \added{qualitatively} analyzed, the paper also considers smart distribution by giving the rationale behind its implementation on the cloud. \cite{Maheshwari2013}\added{ qualitatively} describes a PMU-based state-estimation application and explains how it maps to a cloud architecture. Experimental results show the latency of the system \added{sending data to different locations across the globe.}
A more recent paper \cite{Chai2015} introduces the concept of a publish-subscribe infrastructure\added{ for the smart grid} which is cloud-based.

\replaced{T}{However, t}he sole use of a cloud architecture fulfills only some of the above-mentioned requirements. Using an Internet of Things\added{ (IoT)} \cite{Atzori2014} middleware which is also cloud-based \cite{Gubbi2013} fills the remaining gap. Through resource virtualization, 
which is a common trait of recent \added{IoT} architectural solutions, such a middleware allows for addressing some key properties in 
smart grids, such as: interoperability among various devices in the network, their future-proofness, context-awareness and increased 
capabilities. 

In this paper, an IoT-compliant cloud-based state estimation system for distribution networks in which PMU measurements are virtualized from physical objects is presented. In the considered system only application level processes are put in the cloud, whereas virtualization of physical objects such as PMUs are located at the edge of the communication network (i.e. \deleted{closer to}\added{in the same subnetwork as} the physical object) in order to implement a certain degree of local logic which allows to \deleted{realize} \added{implement} a bandwidth-efficient and smart data transmission to the involved applications in the cloud. 

The major contributions of the paper are the following: 
\begin{itemize}
\item demonstrate that a cloud-based architecture with distributed entities is not only feasible for a \added{flexible and efficient }DSSE application, but also capable of achieving a satisfying Quality of Service (QoS) from both communication and from power systems perspective; 
\item show that part of the logic behind applications can be implemented locally for allowing a certain degree of context-awareness, which prevents cloud applications from bearing unnecessary operations and the communication network from sending unnecessary data.
\end{itemize}

The rest of the paper is organized as follows. The second section is devoted to provide a background about the cloud IoT systems. The third section presents the proposed architectural solution. The fourth and the fifth sections present the considered test system and the experimental results. Then last section draws final conclusions.

\section{Background on cloud IoT systems}
The IoT paradigm has been evolving towards the creation of a cyber-physical world where everything can be found, activated and disactivated, discovered, linked, and updated, so that any possible interaction, involving both virtual and/or physical entities, can take place. Crucial concept of this paradigm is that of the virtual object (VO), which is the digital counterpart of any real entity in the IoT. It has now become a major component of the current IoT platforms, since it is the building block supporting the discovery and aggregation of multiple services, which fosters the creation of complex applications, and addressing heterogeneity and scalability issues. Indeed, virtualization has the ability to: make heterogeneous objects interoperable through the use of semantic descriptions; enable them to acquire, analyse and interpret information about their context in order to \deleted{take}\added{make} relevant decisions and act upon the virtual objects. Moreover, it enhances existing functionalities in the IoT, promoting the creation of new addressing schemes, improving the objects mobility management efficiency, as well as addressing accounting and authentication issues. 

Additionally, IoT platforms are more and more deployed in the cloud, as this approach allows for further boosting some of the IoT  properties such as reliability, always-on availability, elastic processing and memory resource provisioning. These features combined with the previously mentioned ones, make the virtualization and cloud computing the vital technologies for the future IoT solutions. In the rest of the paper, we consider an IoT platform that has been implemented following the mentioned principles, which is named Lysis (http://developers.lysis-iot.com/) \cite{Girau2013}.

\section{Proposed architecture}
The proposed architecture exploits both the virtualization and cloud computing technologies towards a distributed context-aware state estimation. The major characterisitics which are presented in the following subsections are: virtualization of the PMUs, context-awareness at the PMUs, and cloud-IoT based DSSE application.

\begin{figure}[t!]
\centering
\includegraphics [width=0.8\columnwidth] {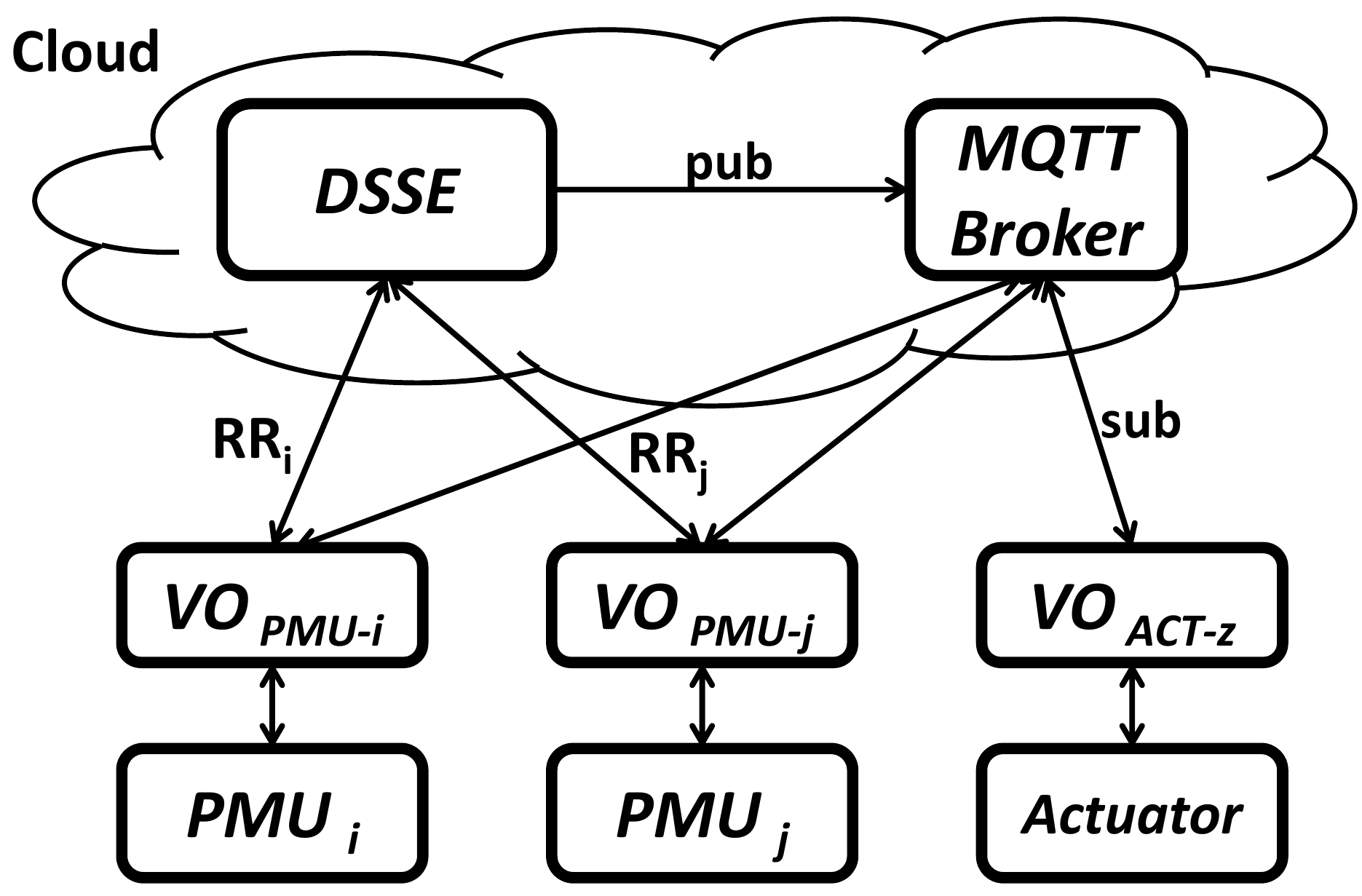}
\caption{System Overview}
\label{syst_overview}
\end{figure}

\subsection{Virtualization of the PMUs}
Fig. \ref{syst_overview} shows a picture of the proposed DSSE system which leverages on the previously mentioned Lysis platform\added{, used for the creation of VOs and the deployment in which they are instatiated and linked to the physical object}. The proposed system is composed of a series of PMUs across a distribution network. These PMUs collect physical quantities such as voltage, current and frequency at a given sampling frequency, which typically can be as high as 50-60 fps. \added{The sampling frequency of the physical PMU cannot be changed at runtime but it is fixed and must be set prior to transmission start.} Gathered measurements are sent with a 
GPS-synchronized timestamp to Virtual Objects (VOs), which are virtual containers implemented at the edge of the communication network 
(i.e. \deleted{closer to}\added{in the same subnetwork as} the physical object)\added{ and receive data from one or more physical 
objects}. A VO is defined as an entity that virtualizes PMU capabilities, so that any application can access or request its resources and 
functionalities in a reusable way, without knowing about the means that are needed to physically reach and retrieve information from the 
physical object. 
In our case, physical objects create a socket with the VO and send measured data using the PMU standard\added{s. 
In particular, the $IEEE\ C.37.118.2$ \cite{IEEE_Std_C37.118.2-2011} for most recent devices compliant
with the $IEEE\ C.37.118.1-2011$ \cite{IEEE_Std_C37.118.1-2011} synchrophasor standard is considered.}  Once VOs receive data from the PMUs, they perform 
relevant processing in order to extract information from the raw measurement data. VOs act as interfaces to the cloud, where PMU gathered 
data are sent at a given Reporting Rate ($RR$). Thanks to the data extraction capabilities, only necessary information can be reported. 
Communication between the cloud and the edge is accomplished using REST APIs and the JSON format for data encapsulation. Therefore data 
are abstracted from the specific PMU format and put in a more convenient and interoperable key-value format. Communications can either 
take place in $GET$ or $PUSH$ mode, which means that data can either be asked to the VO with a HTTP GET query\added{ (useful for infrequent and unpredictable requests)} or be sent automatically to a given location through HTTP POSTs\added{ (useful for example for protection applications requiring periodic updates)}. In the latter case, an appropriate trigger is pre-set in the VO for data forwarding. In 
this paper, on the basis of the specific characteristics of the DSSE application considered, we focus on the case of HTTP POST data forwarding.


\subsection{Context-awareness at the virtual PMUs}
In the past, distribution networks were considered to possess a near-steady state behaviour, which means that measurements do not vary significantly over time. Therefore, information about the network could be sent at a smaller rate than the sensing device (e.g. PMU) maximum, in order to save bandwidth and computation costs. 
Nevertheless, due to the increasing presence of sources of \deleted{uncertainty}\added{variability and unpredictability} such as Distributed Energy Resources (DERs), this assumption is not always valid in today networks. This means that theoretically, PMU measurements should be sent at a sufficient rate to have an up-to-date DSSE which describes the dynamic of the system with a correct resolution. This poses two challenges. First of all, great computation flexibility is necessary, since estimations could pass from one every few seconds to one every signal cycle ($20\ ms$ at a nominal frequency of $50\,\textrm{fps}$). Secondly, sending, for example, $50$ measurements per second of at least $70\ bytes$ of just payload for an unnecessarily long amount of time could easily become unbearable both for the communication network and for the data storage. For this reason, each VO is capable of understanding the meaning of the measured quantities and is able to change its $RR$ according to the needs of the specific use case. Moreover, using a VO, it is also possible to filter unnecessary data thus further decreasing the load which is sent on the channel in critical situations.

\subsection{Cloud-based DSSE application}
Data sent from VOs are received from the DSSE application in the cloud. The DSSE application is responsible for collecting real-time 
measurement data, forecast and hystorical information on the loads and generators \added{(the so-called pseudomeasurements)}, and 
for computing 
the state of the network in terms of both voltage and current phasors. In the following section, in particular, a Weighted Least Square 
(WLS) formulation is adopted.

The DSSE application is implemented using Google App Engines, which provide the necessary computational and storage elasticity required by DSSE. When a new data is received by the application, a new state estimation is triggered for the timestamp indicated in the received packet. Depending on the outcome of the state estimation, countermeasures may be taken if an anomalous event requiring intervention is revealed. In order to do so, the application can communicate in a one-to-one or one-to-many fashion. In the former case, the application exploits the HTTP request of the VO and sends back information as a response to the POST request. In the latter case, the DSSE application block communicates with an MQ Telemetry Transport (MQTT) broker in the cloud. MQTT \cite{mqtt} is a lightweight publish/subscribe protocol in which clients can publish and subscribe messages to a number of topics by means of a broker. This allows DSSE applications to send a single message on a 
given topic related to a specific action involving multiple VOs (e.g. disconnect all generators in a given part of the distribution network). Also actuators are interfaced to the application level by means of a VO, in order to abstract functionalities and give a uniform interface for operational management. 

\section{Application of the Proposed System to the PMU-based DSSE test scenario}

\begin{figure}[t!]
\centering
\includegraphics [width=1\columnwidth, trim=0 50 0 50] {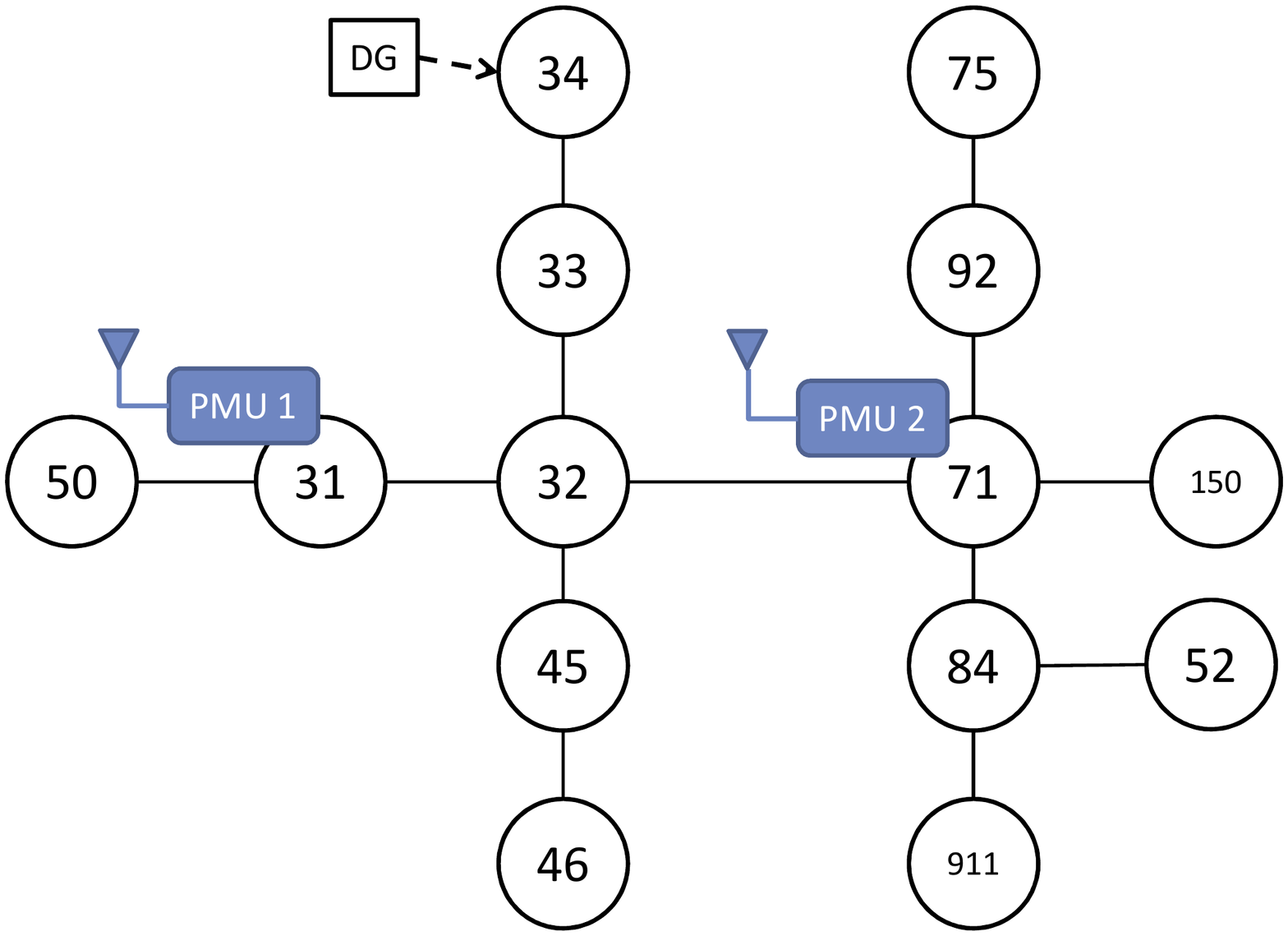}
\caption{Test System}
\label{test_sys}
\end{figure}

The considered \deleted{electrical}\added{electric} system is the sample distribution network \added{derived from} IEEE 13-bus, which is a bus 
radial distribution test feeder \cite{TestSys-IEEE} proposed as a benchmark for the analysis of harmonic propagation in unbalanced networks. For the purposes of this study, the topology and the loads of this network were considered as a starting point to design a test network suitable for the proposed architecture. In particular, the grid used for the test is a totally balanced version of the IEEE 13 bus (Fig. \ref{test_sys}). In addition, a distributed generator was collocated on the network, so that possible transient operating conditions 
 \added{can occur}. The implementation of the network was carried out with the PSCAD/EMTDC software, by Manitoba HVDC. PSCAD is a 
well 
known design and simulation tool to model power systems, acting as a graphical user interface to the EMTDC simulation engine.

A measurement system based on PMUs is supposed. In particular, for a realistic measurement scenario, two PMUs were placed in two points of 
common coupling of the network. One of them is supposed on the root bus, node 31, the other one on the node 71. PMU devices are emulated 
by two real PMU prototypes \added{implemented using a modular platform NI-cRIO and} 
designed to be synchronized by means of GPS receivers and modified to compute synchronized measurements \added{by means of the 
algorithm in \cite{15TIM_DistrPMU}} on the 
pre-stored test signals obtained by PSCAD simulations. In particular, node voltage signals at nodes 31 and 71 are thus considered to be 
measured and the corresponding synchrophasors are computed, along with 
frequency and ROCOF, to give a full set of PMU measurement. 

In the considered scenario, each PMU sends its measurements, 
encapsulated in IEEE C37.118.2 messages to the corresponding VO\added{ (directly connected)} every $20\,\textrm{ms}$, that is the maximum reporting rate indicated by 
IEEE C37.118.1 \cite{IEEE_Std_C37.118.1-2011} standard. In particular, the VO is able to adapt its output reporting rate towards the application, following a fixed policy for all the 
VOs in the network. The VO monitors the incoming RMS voltage values and changes the subsampling rate as follows:
\begin{itemize}
\item if the RMS voltage variation between two consecutive input measurements is above a given threshold ($2 \%$ in the following test results, to account also for PMU accuracy under dynamic conditions), the output $RR_i$ (which is $1\ fps$ in steady-state conditions) is raised up to the maximum $RR$ of $50\ fps$.
\item if the RMS voltage variation between two consecutive output measurements is below a given threshold the output reporting rate is 
decreased by steps (i.e. $25$, $10$ and $1$) at the occurrences of seconds. A threshold value of $0.1 \%$ was used for the tests, 
considering that a commercial PMU is characterized by a very high amplitude measurement accuracy under steady state conditions.
\end{itemize}
The described policy is an example to show the potentiality of the proposed architecture and is chosen considering typical PMU measurement 
accuracies in presence of steady-state and dynamic conditions. 

\added{The DSSE function is performed at the application level using the measurements received from different VOs at varying $RR$. 
In fact, the measurements (with the corresponding timestamps), originated by different PMUs can, depending on the events detected locally, 
reach the DSSE at different rates. For this reason, the DSSE function is performed coordinating the different measurement flows coming 
from VOs. The most recent measurements are considered at each estimation and corresponding timestamps are aligned. As an example, a higher 
VO rate triggers a higher DSSE update rate, thus allowing to follow more accurately a faster event. The state estimation is particularly 
useful to track the behaviour even at nodes that are not equipped with instrumentation. 
The performed DSSE relies on the knowledge of the network model and is implemented by means of the fast 
WLS branch-current state 
technique presented in \cite{PauPegSul13TIM_EffBranchCurrentDSSE}, exploiting the linearization of power injection 
pseudomeasurements (obtained from load profiles) to obtain a constant Gain matrix in WLS computation.}

In the considered scenario, for the sake of comparison, DSSE can be done in two different ways:
\begin{enumerate}
\item all VOs send their measurements at the fixed maximum rate so that state estimation is computed in the most fine-grained manner; 
\item measurement $RR$s are tuned according to the algorithm previously described.  
\end{enumerate}
   
The rationale behind testing these two scenarios is to demonstrate that using a VO which implements a local logic allows to tune the measurement $RR$ so that sufficient information on the network dynamics is kept while reducing, at the same time, the load of packets sent to the cloud. Moreover, the latency associated with various $RR$ is discussed.

In order to test the dynamics of the measurement system, several events have been generated and measured on the network. Here, for the 
sake of simplicity, the most significant concerning the insertion of a distributed generator is reported. The event is simulated in a 
given moment preceded by a steady-state of the network in which $RR=1\,\textrm{fps}$. PMUs are continuously monitoring the operating conditions of the 
grid and measurements are collected by the VOs. Therefore, the variations of the RMS values of the measured voltage are monitored and in 
case such variations exceed a certain threshold then a thicker monitoring is prompted.

\section{Results}

\begin{figure}[t!]
\centering
\includegraphics [width=1\columnwidth] {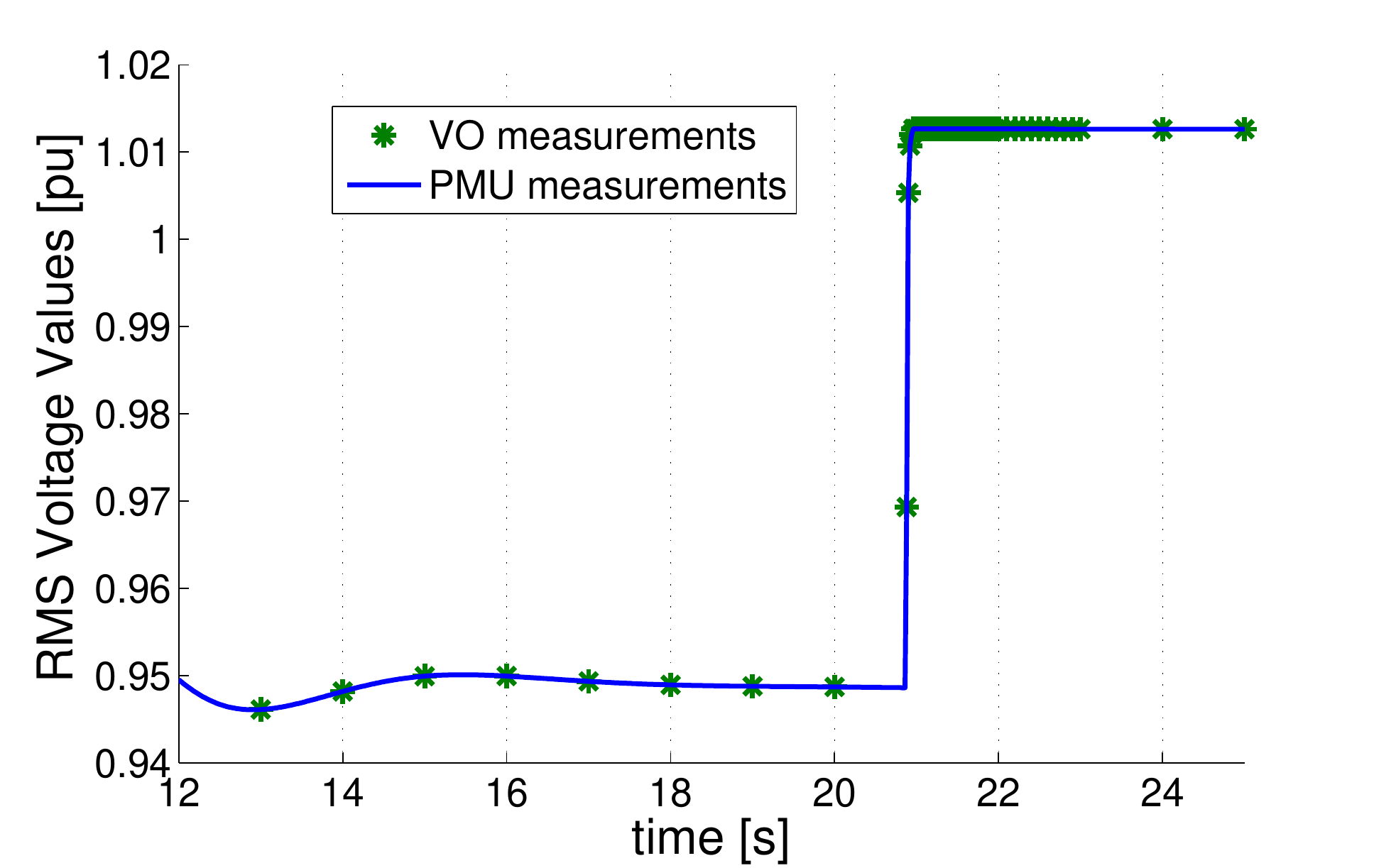}
\caption{PMU measurements at 50 fps and values sent by the VO to the DSSE application in the cloud for the directly monitored node 31}
\label{node1}
\end{figure}

Fig. \ref{node1} shows the measurements gathered by the PMU at node 31 (a continuous line is used, for the sake of clarity, to connect the thin grid points obtained on the 20 ms resolution scale), compared with those transmitted by the VO to the DSSE application. It can be noticed how the actual $RR$ clearly follows the evolution of the RMS voltage and the VO correctly detects the fast transient, thus triggering a faster reporting rate which suceeds in following the dynamic of the system. On the other hand, the DSSE application takes more frequent snapshots of the network status only when an interesting phenomenon need to be investigated.

\begin{figure}[t!]
\centering
\includegraphics [width=1\columnwidth] {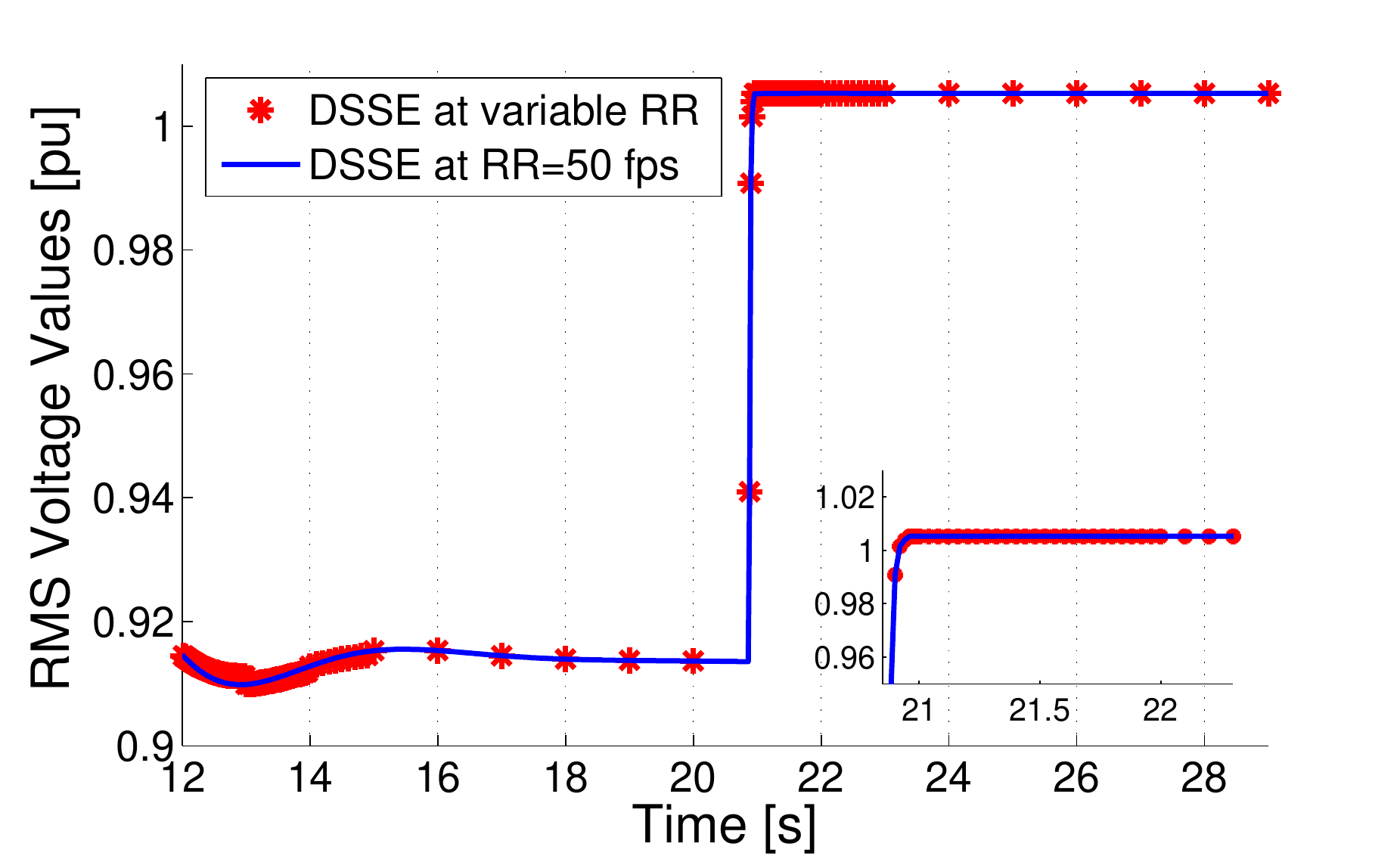}
\caption{DSSE for node 33 which is not PMU-monitored. The zoomed inset highlights the abrupt transition instant and the decrease of RR at 
the $22^{nd}$ second, which is not visible in the bigger plot.}
\label{node2}
\end{figure}

Concerning the state estimation, Fig. \ref{node2} shows the results of the DSSE in terms of the RMS voltage estimation at node 33 (not 
PMU-monitored) when both the maximum fixed $RR=50\,\textrm{fps}$ given by the PMUs and the selection of the rate is used. In the 
considered scenario, after an initial transient state and a subsequent stabilization, the insertion of the DER \added{at $20.9\ s$} leads to a new increase of 
node voltage amplitude \added{(because of the inversely injected power flow)}. At this point the DSSE promptly reacts to the dynamic 
of the system and follows the dynamic thanks to the adaptive policy of VOs and the variable DSSE reporting rate.

\begin{figure}[t!]
\centering
\includegraphics [width=1\columnwidth] {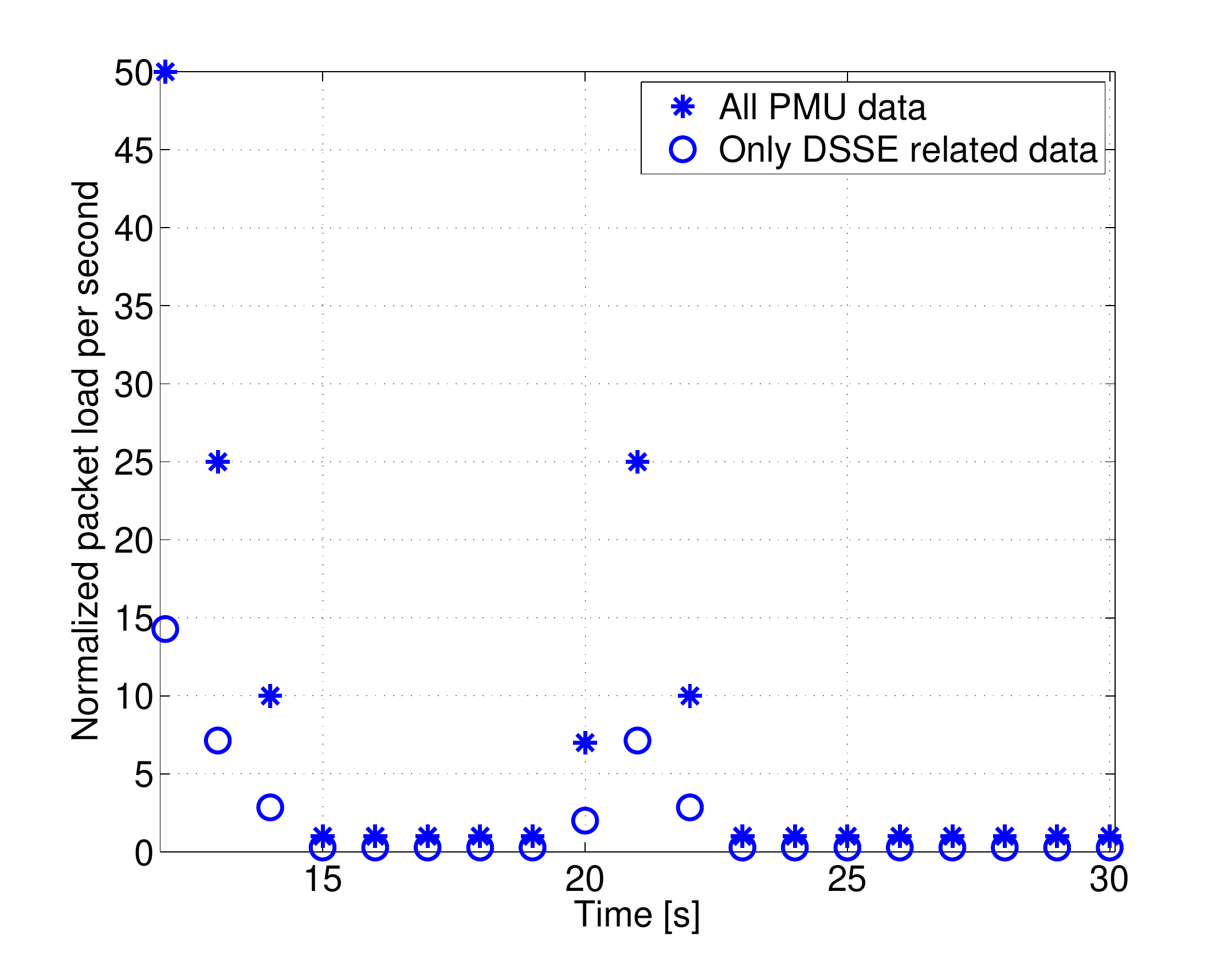}
\caption{Network load generated by each PMU normalized over the number of PMUs.}
\label{load}
\end{figure}

Fig. \ref{load} shows the traffic load related to the adaptive $RR$ of the PMU, normalized over the packet size and the number of PMUs. For the case of DSSE operating at a fixed rate of $RR=50\,\textrm{fps}$, the best state estimation is obtained. However, a huge amount of redundant and unuseful data is sent which is most likely to occupy storage space in the cloud and overload the communication network. 
Considering our specific case, $2$ PMUs sending $70$ Bytes long messages (which is the lowest possible length) at $50\,\textrm{fps}$ generate $25.2$ MB per hour and almost $605$ MB in a whole day. While this amount is surely affordable for a cloud infrastructure, it is straightforward to think that this approach becomes rapidly impractical with big distribution networks in which several PMUs will be placed for state estimation.

With the proposed approach instead, the rate of the state estimation is generally low (in this case $1\,\textrm{fps}$) unless 
consistent dynamics take place, as shown in Fig. \ref{load} for our reference testing scenario. We want to highlight here, that our insights based on the test scenario can easily be considered as a worst case, since the portion of time window in which a dynamic is present is much bigger than what is found in the average behaviour of a distribution network over an extended period. Nevertheless, it gives a rough idea of the data savings. The average load on the channel in the case of a variable $RR$ is less than $15\,\textrm{fps}$, which means $3.7$ MB of data each hour and about $89$ MB of data per day. 

But this is not the best performance that our system can achieve. As a matter of fact, having a VO which implements some kind of local intelligence, allows the system to filter data depending on the interest of every application. This is one of the advantages of having local entities which abstract physical object properties. In this last case, considering that state estimation only requires timestamps and phasors measurements, the quantity of data sent can be reduced of about the 70\% with respect to the case in which all the information in the PMU data packet is sent. Therefore, referring to the considered scenario, the load generated will be $1$ MB per hour and about $25$ MB for the whole day.

Even though these are estimated results calculated in a worst case, they give an idea of how much can be saved through the implementation of a local VO having context-awareness capabilities.


Finally, let us discuss the latency of the considered scenario. Table \ref{latency_table} shows the average latency of the overall system considering that a measurement packet is sent to the cloud application and a certain message is passed back to the VO. The delay considered includes: 
\begin{itemize}
\item the round trip time on the communication network, which in our case is the path between the subnetwork of the University of Cagliari and the Google farm where app engines are hosted, which is located in southern Belgium;
\item the time needed in order to accomplish DSSE state estimation;
\item the time needed to store data of the accomplished DSSE state estimation. 
\end{itemize}

In this particular case, 5 \added{Google App Engine} instances of class B8 (which means $4.8\ Ghz$ processor) have been used. As we can see, the use of a cloud infrastructure allows to manage different $RR$ in an elastic manner. As a matter of fact, the computed average delay only slightly degrades when increasing the RR. Moreover, for any RR, the average delay is between $257$ and $267\ ms$.

Table \ref{performance} shows the different performance classes defined by $IEC 61850$ for smart grid management and can be used here as a 
useful meter of comparison for the latency obtained in our system. It can be seen that the proposed system satisfies the requirements of 
classes $TT0$, $TT1$ and $TT2$. In particular, performance classes $TT0$ and $TT1$ have $100\%$ dependability \added{(which means that the 
delay of any packet remains under the $1000\ ms$ threshold) }while for class $TT2$ the dependability slightly decreases as the RR 
increases. Such a result is very promising for the future of power systems.

In fact, performance of most functions in energy management systems depend on the DSSE results. Monitoring functions can issue warnings and recommendation for load shedding with rates in the order of the second so that remedial action scheme is applied. The inertia of the grids commonly permits slow voltage fluctuations. As a consequence, in the most advanced applications proposed by Transmission System Operators and Distribution System Operators voltage regulation can be done requiring adjustments in the order of seconds. Therefore, the obtained results prove that advanced monitoring functions can be performed on the basis of the proposed architecture.

\begin{table}[t!]
\centering
\includegraphics [width=0.8\columnwidth] {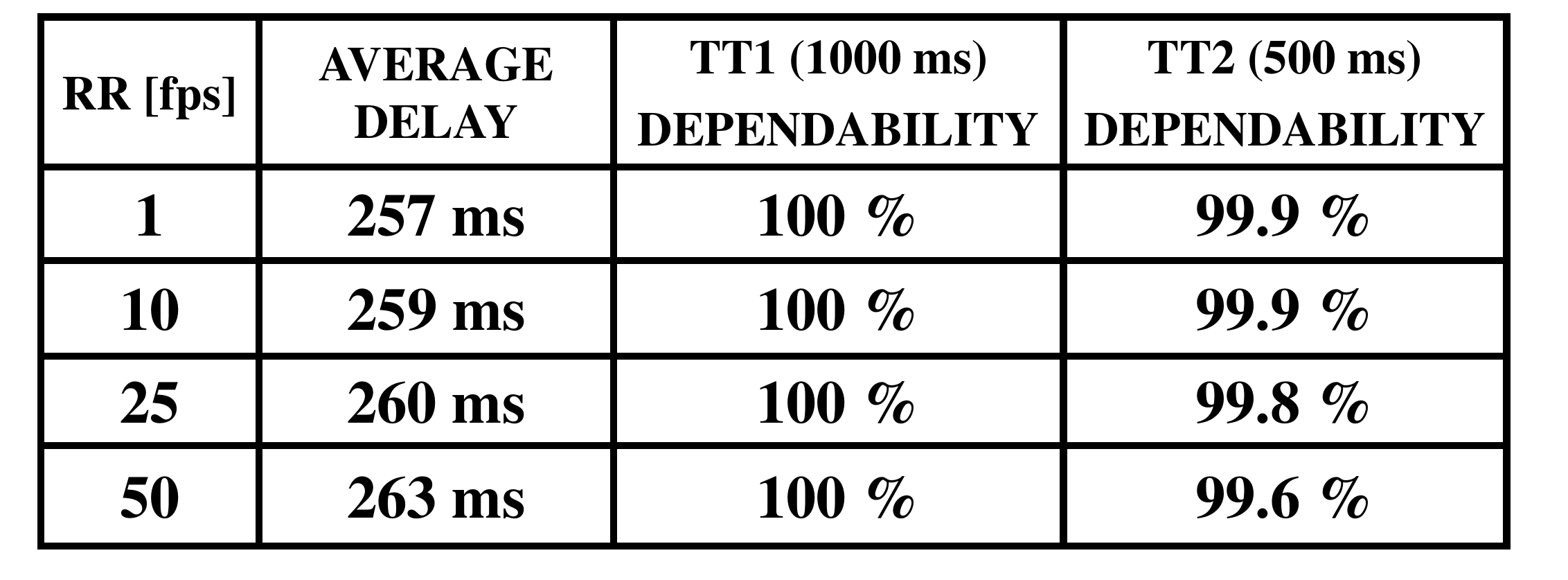}
\caption{Average delay and dependability for TT1 and TT2 performance classes.}
\label{latency_table}
\end{table}

\begin{table}[t!]
\centering
\includegraphics [width=0.97\columnwidth] {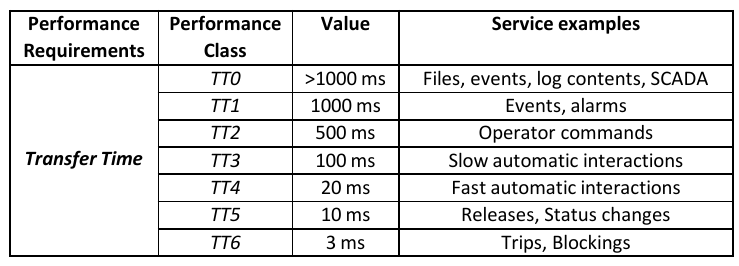}
\caption{Communication requirements for power systems \cite{iec}}
\label{performance}
\end{table}

\section{Conclusions and Future Work}
In this paper a solution for the implementation of a distributed system state estimation in an IoT cloud-based system has been proposed. The major features are: the virtualization of the PMUs with the virtual objects running on the edge of the communications network and the implementation of a logic at the virtual object so that the transmission rate of the measurements is adapted on the basis of the foreseen usefulness for the state estimation accuracy. The performance analysis conducted in a distributed \deleted{electrical}\added{electric} network derived from the IEEE 13-bus has shown that the required QoS application-specific requirements can be assured even with a reduction in the transmission rate of over 95\% compared with the case of full-operating PMUs. Moreover, the found latencies have been demonstrated to be compliant with the requirements of the domain considered.
As a future work, we intend to evaluate the load and the latency in a bigger network in which a greater number of PMUs is present. In fact, the resulting load and latency may not be tolerated by all the possible applications, resulting in more stringent requirements to be imposed. Moreover, some more studies will be done to further optimize the latency of the system.  Finally, we intend to evaluate the usefulness of the proposed IoT-cloud system for other smart grid applications and the possibility to implement more state estimation logic at the VO, in order to satisfy latency constraints of stricter performance classes.

\bibliographystyle{IEEEtran}
\bibliography{IEEEabrv,references,stateestimationandpowerflow,synchrophasors,meterplacement}

\end{document}